\documentclass[twocolumn,showpacs,preprintnumbers,amsmath,amssymb]{revtex4}
\usepackage{graphicx}% Include figure files
\usepackage{dcolumn}% Align table columns on decimal point
\usepackage{bm}% bold math

\usepackage[pagebackref=true]{hyperref}
\hypersetup{pdftitle = The title of my PDF, pdfauthor = My name, pdfsubject= The subject, pdfkeywords = keyword1 keyword2 keyword3}
\hypersetup{colorlinks = true, linkcolor = green, anchorcolor = red, citecolor = blue, filecolor = red, pagecolor = red, urlcolor = red}

\newcommand{\pasa}{Publications of the Astronomical Society of Australia}
\newcommand{\aap}{Astronom. and Astrophys.}
\newcommand{\aaps}{Astronom. and Astrophys. Suppl. Ser.}
\newcommand{\aj}{Astronom. J.}
\newcommand{\apjs}{Astrophys. J. Suppl.}
\newcommand{\mnras}{Monthly Notices Roy. Astronom. Soc.}
\def\farcs{\hbox{$.\!\!^{\prime\prime}$}}
\def\fm{\hbox{$.\!\!^m$}}
\def\fs{\hbox{$.\!\!^s$}}
\def\degr{\hbox{$^\circ$}}

\begin{document}

%\titlerunning{PARAMETERS OF THE  VISUALLY CLOSE BINARY SYSTEM HIP11253 (HD14874)}
%\authorrunning{AL-WARDAT and WIDYAN}%
%\toctitle{Parameters of the  visually close binary system Hip11253 (HD14874)}
%\tocauthor{}%
\title{Physical and Geometrical Parameters of the Evolved Binary System HD6009}
\author{\firstname{M.\,A.}~\surname{Al-Wardat}}
\email{mwardat@ahu.edu.jo}
\affiliation{Department of Physics, Al-Hussein Bin Talal University, P.O.Box 20, 71111, Ma'an, Jordan.}
%\author{\firstname{Yu.\, Yu.}~\surname{ Balega}}
%\email{balega@sao.ru}
%\affiliation{Special Astrophysical Observatory of the Russian AS, Nizhnij Arkhyz, 369167 Russia}
%\author{\firstname{R.\, Ya.}~\surname{ Zhuchkov}}
%\affiliation{Department of Astronomy and Geodezy, Kazan Federal Uni., Kremlevskaya str. 18, Kazan, 420008, Russia}

%\received{ , 2013}%
\received{November 29, 2013}%
\revised{2014}%

%   \date{\today}
\date{August 05, 2014}

\begin{abstract}
  Atmospheric modeling and dynamical analysis of the components of the visually close  binary
system (VCBS) HD6009 were used to estimate their individual physical and geometrical parameters. Model atmospheres  were constructed using a grid of  Kurucz solar metalicity blanketed models, and used to compute the individual synthetic spectral energy distribution (SED) for each component separately. These SEDs were combined together to compute the entire SED for the system from the net luminosities of the components $a$ and $b$ located at a distance $d$ from the Earth.
%The entire observational SED of the system was used as a reference for the comparison with the synthetic ones.
 We used the feedback modified parameters and iteration method to get the best fit between synthetic and  observational entire SEDs.

 The physical and geometrical parameters of the system's components were derived as: $T_{\rm eff}^{a}
=5625\pm75$\,K, $T_{\rm eff}^{b} =5575\pm75$\,K, log $g_{a}=3.75\pm0.25$,
 log $g_{b}=3.75\pm0.25$, $R_{a}=2.75\pm0.30 R_\odot$,  $R_{b}=2.65\pm0.30 R_\odot$, $M_v^{a}= 2\fm80\pm0.30$, $M_v^{b}=2\fm93\pm0.30$, $M_a= 1.42\pm0.15 M_{\odot}$,  $M_b=1.40\pm0.15 M_{\odot}$, $L_a=6.80\pm0.75 L_\odot$,   $L_b=6.09\pm0.75 L_\odot$ and $\pi=14.43$mas dynamical parallax.
 The system is shown to be consist of G6 IV primary and G6 IV secondary components.

\end{abstract}

\pacs{95.75.Fg, 97.10.Ex, 97.10.Pg, 97.10.Ri, 97.20.Jg, 97.80.Fk}
%\keywords{stars: binaries: visual; stars: interferometric binaries; stars: photometry}

\maketitle

\section{Introduction}
Hipparcos mission revealed that many of the previously known as  single stars are actually binary or multiple systems, rasing by that the duplicity and multiplicity of galactic stellar systems. Most  of these binary and multiple systems are nearby stars, which appears as a single star even with largest ground based telescopes except by means of  modern high resolution techniques like speckle interferometry  (SI) and adaptive optics (AO). These systems are known as visually close (spatially unresolved on the sky) binary systems (VCBS).

The study of binary systems, in general, plays an important role in determining several key stellar parameters, which is more complicated in the case of the evolved VCBS.
In spite the fact that hundreds of binary systems with periods in the order of 10 years or less, are routinely observed by different
groups of high resolution techniques  around the world, there is still a lake in the determination of  the individual physical parameters of the systems'
components. So, the combination of speckle interferometry, spectrophotometry, atmospheres modelling and -recently- orbital analysis opens a new window in the accurate determination of the physical and geometrical parameters of VCBS. These parameters include the effective temperatures, radii,
orbital elements, spectral types, luminosity classes and masses for both components of the binary
system. The method was successfully applied to several main sequence binary systems: ADS11061, Cou1289, Cou1291, Hip11352,  Hip11253, Hip70973 and Hip72479  \citep{2002BSAO...53...51A, 2007AN....328...63A, 2009AN....330..385A, 2009AstBu..64..365A, 2012PASA...29..523A}, and to the subgiant systems HD25811 and HD375 \citep{2013arXiv1311.5721A, 2013arXiv1311.5737A}.

 The system HD6009 (Hip4809) is a well known VCB, which is routinely observed by different SI groups around the world. It was first proposed as an evolved G9 star with luminosity class between III and IV by \cite{1961ApJ...134..809Y}  from his analysis of the strength of the $\lambda$4077 ionized strontium line based on the prism spectra from the Curtis Schmidt telescope of the University of Michigan and slit spectrograms from the 60 inch reflector of the Mt. Wilson Observatory, which was  supported as IV by the moderate cyanogen absorption at  $\lambda$4216. \cite{2002aaa...385...87B} also  noted that the computed absolute magnitudes of the individual components as $M_{va}=3.1 \& M_{vb}=3.3$  with their late G spectral type do not correspond to their ($B-V$)or ($V-I$) colours, which means that the system is evolved. Moreover, \cite{2006aaa...448..703B}  found that the system consists of G6 and G9 subgiant stars depending on the magnitude difference and Hipparcos parallax measurements.

 So, the analysis of the system the aforementioned method would help  in  understanding the formation and evolution mechanisms of  stellar binary systems.

 Table  ~\ref{table1}contains the SIMBAD data of the system, and Table  ~\ref{table2}
contains data from Hipparcos and Tycho Catalogues \citep{1997yCat.1239....0E}.

\begin{table}
\begin{center}
\caption{Data from SIMBAD} \label{table1}
\begin{tabular}{l|c}\hline
  & HD6009 \\
  & Hip4809    \\
 \hline
$\alpha_{2000}$ & $01^h 01^m43\fs581$    \\
$\delta_{2000}$&$+25\degr17'31.''99$ \\
 Tyc &  1743-1174-1 \\
 HD &  6009  \\
 Spectral type & G8IV  \\
\hline
\end{tabular}
\end{center}
\end{table}

\begin{table}
\begin{center}
\caption{Data from Hipparcos and Tycho Catalogues} \label{table2}
\begin{tabular}{l|c}\hline
  & HD6009 \\
  & Hip4809    \\
 \hline
  $V_J(Hip)$ & $6\fm71$  \\
  $B_T$ & $7\fm683\pm0.008$  \\
 $V_T$ & $6\fm806\pm0.006$  \\
 $(B-V)_J(Tycho)$ & $0\fm783\pm0.007$  \\
 Hip Trig. Parallax (mas) & $13.94\pm0.90$ \\
 Tyc Trig. Parallax (mas) & $12.3\pm5.3$ \\
\hline
\end{tabular}
\end{center}
\end{table}

 Table~\ref{tableresult} lists the synthetic  Jonson, Str\"{o}mgren and Tycho magnitudes and color indices, which were calculated by \cite{2002BSAO...53...58A, 2008AstBu..63..361A}, depending on the  observational  SED.

\begin{table}[!ht]
\begin{center}
\caption{Jonson, Str\"{o}mgren and Tycho  magnitudes and  color indices of the entire system \cite{2002BSAO...53...58A, 2008AstBu..63..361A}.}
\label{tableresult}
\begin{tabular}{l|c}
\noalign{\smallskip}
\hline
\noalign{\smallskip}
  & HD6009 \\
  & Hip4809   \\
 \hline
\noalign{\smallskip}
 $B_J$ & $7\fm43 \pm 0.06  $ \\
 $R_J$ & $6\fm32 \pm 0.07 $  \\
 $V_J$ & $6\fm74 \pm 0.06 $ \\
 $(B-V)_J$ & $0\fm67 \pm 0.08$ \\
 $v$ & $7\fm85 \pm 0.06$ \\
 $b$& $7\fm145 \pm 0.06$ \\
 $y$ & $6\fm70 \pm 0.06$ \\
$v-b$& $0\fm72 \pm 0.08$ \\
 $b-y$ & $0\fm44 \pm 0.08$ \\
  $B_T$ & $7\fm64 \pm 0.06  $ \\
 $V_T$ & $6\fm81 \pm 0.06 $ \\
 $(B-V)_T$ & $0\fm82 \pm 0.08$ \\
\noalign{\smallskip}
\hline
\end{tabular}
\end{center}
\end{table}

\section{ORBITAL ELEMENTS}
\label{orbital_elements}

\begin{figure}[!h]
\resizebox{\hsize}{!} {\includegraphics[]{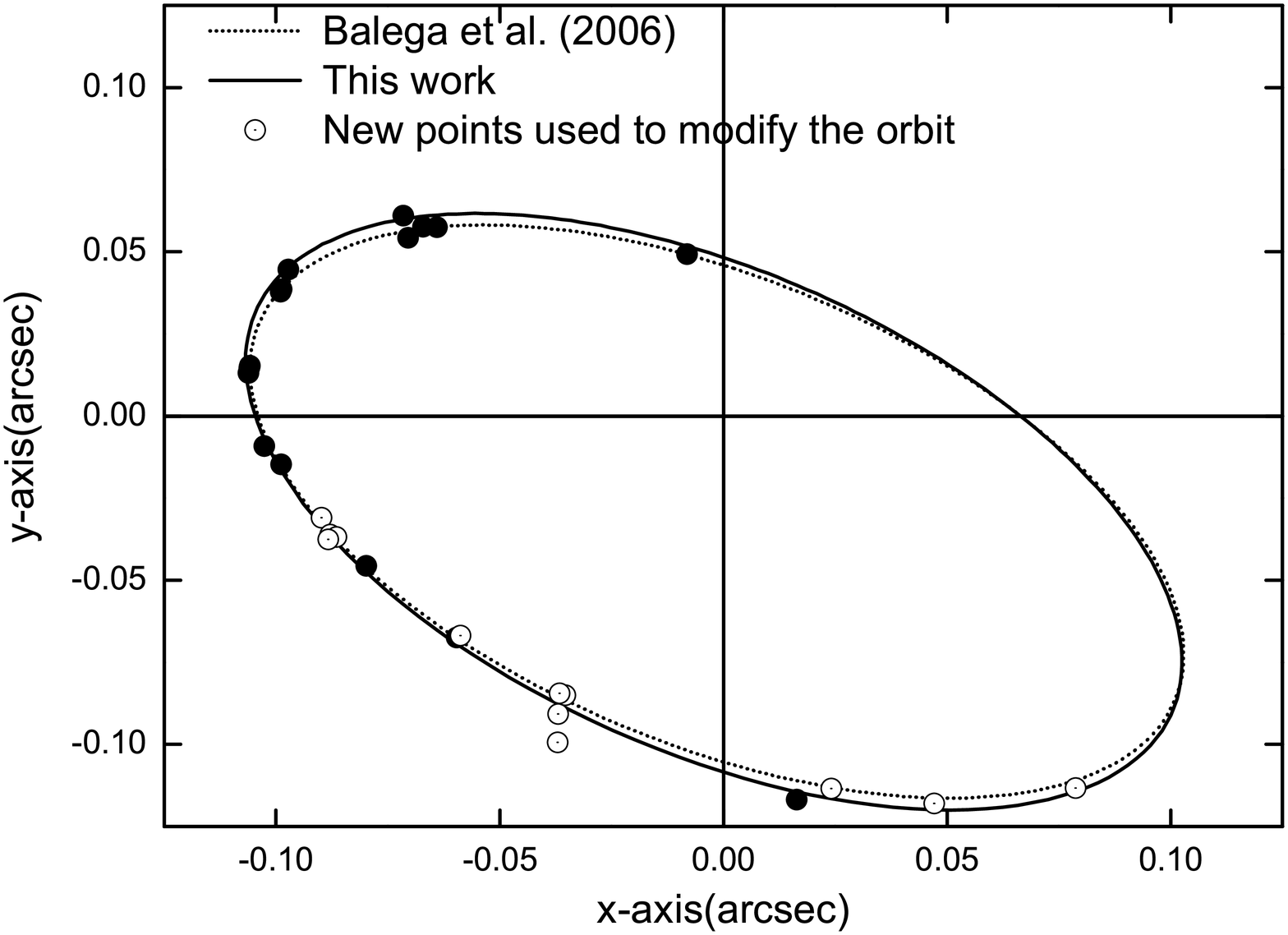}}
%\centerline{\psfig{figure=FigChPP/Hip/ModFitHip.eps,width=8.5\textwidth,clip=}}
 %\centerline{\psfig{figure=Modfit.eps,width=0.70\textwidth,clip=}}%,width=14cm}}
 \caption{The modified relative visual orbit of the system in this work (solid line) against  that of \cite{2006aaa...448..703B} (doted line).}
 \label{orbit_all}
\end{figure}

\cite{2006BSAO...59...20B} introduced  a preliminary orbit of the system depending on 12 interferometric measurements and the first Hipparcos data point with a period of about 15 years. The orbit was then modified by \cite{2006aaa...448..703B} using 16 astrometric measurements (black dots in Fig. ~\ref{orbit_all}).

 A slight modification of the orbit is introduced here using all relative positional measurements listed in Table ~\ref{points}, which includes 12 more measurements than those used by \cite{2006aaa...448..703B} and cover the complete orbit since the first measurement of Hipparcos (Fig. ~\ref{orbit_all}). The quadrants of some of these measurements were adjusted to obtain a consistent orbit.

\begin{table*}
\begin{center}
\caption{Modified orbital elements  of the system with those of the old orbit by  \citep{2006aaa...448..703B}.}
% \small
 \label{orbital_elements}
\begin{tabular}{lccc}
\hline \hline
  Parameters                & This work             &  Old orbit by \citep{2006aaa...448..703B} $\dagger$\\
%check error values

 \hline
 Period $P$                 & $16.28^y\pm 0.08^y$	&$16.41^y\pm 0.11^y$ \\
%   \hline
%\noalign{\smallskip}
 Periastron epoch  $T_0$   &   $1998.55\pm 0.06$   &$1998.62\pm 0.02$   \\
%   \hline
%\noalign{\smallskip}
 Eccentricity $e$           & $   0.385\pm 0.014$   &$0.393\pm 0.012$    \\
%   \hline
%\noalign{\smallskip}
  Semi-major axis $a$       & $   0\farcs130 \pm 0.002$ &$0\farcs128\pm 0.002$ \\
%    \hline
%\noalign{\smallskip}
  Inclination $i$           & $58.18^\circ \pm 0.93$  &$	58.4^\circ \pm 0.5 $    \\
%   \hline
%\noalign{\smallskip}
 Argument of periastron $\omega$    &$287.45^\circ \pm 0.60$  &$	106.9^\circ \pm 0.6 $    \\
%   \hline
%\noalign{\smallskip}
 Position angle of nodes $\Omega$   & $235.03^\circ \pm 0.68$    &$	57.1^\circ \pm 0.8$     \\
   \hline \hline
%\noalign{\smallskip}
\end{tabular}\\
%$\dagger$\citep{2006aaa...448..703B},
%$\ddagger$\citep{2001AstL...27...95B}.
\end{center}
\end{table*}

The estimated orbital elements  of the system along with those of the old
orbit are listed in Table ~\ref{orbital_elements}.

\begin{figure}[!h]
\resizebox{\hsize}{!} {\includegraphics[]{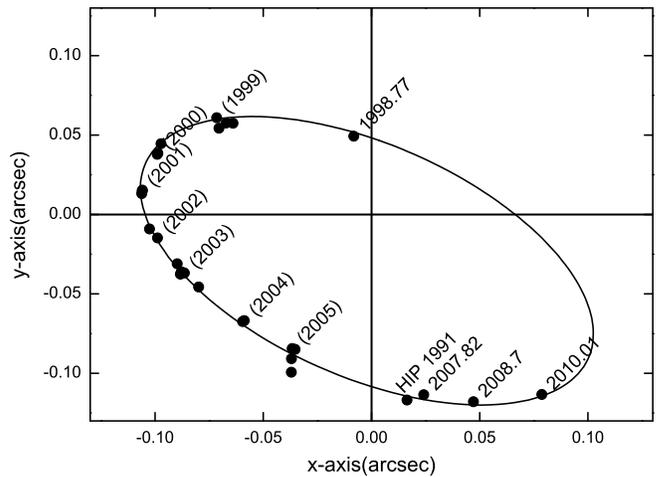}}
%\centerline{\psfig{figure=FigChPP/Hip/ModFitHip.eps,width=8.5\textwidth,clip=}}
 %\centerline{\psfig{figure=Modfit.eps,width=0.70\textwidth,clip=}}%,width=14cm}}
 \caption{Relative visual orbit of the system with the epoch of the positional measurements; the origin represents the position of the primary component.}
 \label{orbit}
\end{figure}

Fig. ~\ref{orbit} shows the relative visual orbit of the system with the epoch of the positional measurements, and Fig. ~\ref{orbit_all} shows the new orbit against the old one of \citep{2006aaa...448..703B}.

%----------------------------------------------------------

\begin{table*}
\begin{center}
\caption{Relative positional measurements using different methods which are used to build
the orbit of the system. These points are taken from the Fourth Catalog of
Interferometric Measurements of Binary Stars. }
\label{points}
\begin{tabular}{lccccccccccc}\hline \hline
  Date     &$\theta$(deg)&$\sigma_\theta$& $\rho$(deg)&$\sigma_\rho$&$\triangle m$&$\sigma_{\triangle m}$&filter($\lambda$)&$\Delta\lambda$)& Tel.  &Ref.$^\dag$ & Meth.$^\ddag$   \\
\hline
1991.25     &  188      &  1.5  &    0.118  &  0.004 &   0.28   &   0.30    &   511 & 222   & 0.3   &  HIP1997a \cite{1997yCat.1239....0E}  & Hh\\
1997.7206   &           &       &           &$<$0.065&          &           &   550 &   24  & 2.1   &   Msn1999b \cite{1999AJ....117.1890M}& Su \\
1998.7718	&	0	    &	0	&	0	    &	0	 &	0.1	    &	0.24	&	545	&	30	&	6	&	Plz2005	\cite{2005aaa...431..587P} &	S\\
1998.7718	&	189.5	&	0.5	&	0.05	&	0.002&	0.17	&	0.15	&	545	&	30	&	6	&	Bag2002	\cite{2002aaa...385...87B} &	S\\
1999.7472	&	228.2	&	1.5	&	0.086	&	0.002&	0.3	    &	0.16	&	2115&	214	&	6	&	Bag2002	\cite{2002aaa...385...87B} &	S\\
1999.8128	&	49.5	&	0.3	&	0.0885	&  0.0003&	0.19	&	0.04	&	610	&	20	&	6	&	Bag2004	\cite{2004aaa...422..627B} &	S\\
1999.8233	&	49.7	&	0	&	0.094	&	0	 &	0	    &	0   	&	550	&	24	&	2.1	&	Msn2001b \cite{2001AJ....121.3224M}&	Su\\
1999.8882	&	0	    &	0	&	0	    &	0	 &	0.05	&	0.15	&	648	&	41	&	3.5	&	Hor2004	\cite{2004AJ....127.1727H} &	S\\
1999.8882	&	52.5	&	0	&	0.089	&	0	 &	0	    &	0   	&	648	&	41	&	3.5	&	Hor2002a \cite{2002AJ....123.3442H}&	S\\
2000.7674	&	0	    &	0	&	0	    &	0	 &	0.22	&	0.15	&	503	&	40	&	3.5	&	Hor2004	\cite{2004AJ....127.1727H} &	S\\
2000.7674	&	65.5	&	0	&	0.107	&	0	 &	0	    &	0   	&	503	&	40	&	3.5	&	Hor2002a \cite{2002AJ....123.3442H}&	S\\
2000.8728	&	249.2	&	0.6	&	0.106	&	0.001&	0.12	&	0.19	&	800	&	110	&	6	&	Bag2006b \cite{2006BSAO...59...20B}&	S\\
2000.8755	&	248.8	&	0.4	&	0.106	&	0.001&	0.16	&	0.05	&	600	&	30	&	6	&	Bag2006b \cite{2006BSAO...59...20B}&	S\\
2001.7526	&	261.9	&	0.2	&	0.107	&	0.001&	0      	&	0.12	&	545	&	30	&	6	&	Bag2006b \cite{2006BSAO...59...20B}&	S\\
2001.7526	&	262.2	&	0.2	&	0.107	&	0.001&	0      	&	0.12	&	600	&	30	&	6	&	Bag2006b \cite{2006BSAO...59...20B}&	S\\
2001.753	&	262.2	&	0	&	0.107	&	0	 &	0      	&	0     	&	--	&	--	&	6	&	Bag2006	\cite{2006aaa...448..703B} &	S\\
2001.845	&	263.1	&	0	&	0.107	&	0	 &	0      	&	0     	&	--	&	--	&	6	&	Bag2006	\cite{2006aaa...448..703B}&	S\\
2002.726	&	275.2	&	0	&	0.103	&	0	 &	0   	&	0    	&	--	&	--	&	6	&	Bag2006	\cite{2006aaa...448..703B}&	S\\
2002.796	&	278.6	&	0	&	0.1	    &	0	 &	0     	&	0    	&	--	&	--	&	6	&	Bag2006	\cite{2006aaa...448..703B}&	S\\
2003.6372	&	109.2	&	0	&	0.095	&	0	 &	0	    &	0   	&	550	&	40	&	3.5	&	Hor2008	\cite{2008AJ....136..312H}&	S\\
2003.6372	&	112.5	&	0	&	0.095	&	0	 &	0.15	&	0   	&	754	&	44	&	3.5	&	Hor2008	\cite{2008AJ....136..312H}&	S\\
2003.6372	&	113.2	&	0	&	0.094	&	0	 &	0      	&	0    	&	698	&	39	&	3.5	&	Hor2008	\cite{2008AJ....136..312H}&	S\\
2003.6372	&	293.2	&	0	&	0.096	&	0	 &	0      	&	0	    &	650	&	38	&	3.5	&	Hor2008	\cite{2008AJ....136..312H}&	S\\
2003.928	&	299.9	&	0	&	0.092	&	0	 &	0      	&	0     	&	--	&	--	&	6	&	Bag2006	\cite{2006aaa...448..703B}&	S\\
2004.815	&	318.7	&	0	&	0.09	&	0	 &	0       &	0      	&	--	&	--	&	6	&	Bag2006	\cite{2006aaa...448..703B}&	S\\
2004.8154	&	318.8	&	0.3	&	0.089	&	0.002&	0.21	&	0.04	&	600	&	30	&	6	&	Bag2007b\cite{2007AstBu..62..339B}&	S\\
2005.5619	&	338	    &	0	&	0.098	&	0	 &	0.81	&	0     	&	698	&	39	&	3.5	&	Hor2008	\cite{2008AJ....136..312H}&	S\\
2005.5975	&	337.6	&	0	&	0.092	&	0	 &	0.06	&	0   	&	698	&	39	&	3.5	&	Hor2008	\cite{2008AJ....136..312H}&	S\\
2005.5975	&	336.7	&	0	&	0.092	&	0	 &	0      	&	0    	&	698	&	39	&	3.5	&	Hor2008	\cite{2008AJ....136..312H}&	S\\
2005.8627	&	339.7	&	0	&	0.106	&	0	 &	0      	&	0	    &	550	&	24	&	3.8	&	Msn2009	\cite{2009AJ....137.3358M}&	Su\\
2007.8228	&	192	    &	0	&	0.116	&	0	 &	0.26	&	0     	&	698	&	39	&	3.5	&	Hor2010	\cite{2010AJ....139..205H}&	S\\
2008.702	&	201.8	&	1.1	&	0.127	&	0.003&	0       &	0.12	&	550	&	40	&	3.5	&	Hor2012a \cite{2012AJ....143...10H}&	S\\
2010.0045	&	214.8	&	0	&	0.138	&	0	 &	0.23	&	0   	&	562	&	40	&	3.5	&	Hor2011	\cite{2011AJ....141...45H}&	S\\
2010.0045	&	0	    &	0	&	0	    &	0	 &	0.15	&	0   	&	692	&	40	&	3.5	&	Hor2011	\cite{2011AJ....141...45H}&	S\\

\hline \hline
\end{tabular}
\end{center}
$^\dag${References are abbreviated as in the Fourth Catalog of Interferometric Measurements of Binary Stars}.\\
$^\ddag${ S: Speckle Interferometry, Su: USNO speckle}.
\end{table*}

\section{Atmospheric modelling}
%\subsection{Observational reference spectrum}

%The  observational spectral energy distributions (SED) of the system was taken from \cite{2002BSAO...53...58A}.
%Note that some of the strong lines and depressions, especially in the red part of the spectrum (around $\lambda 6867\AA$, $\lambda 7200\AA$, and $\lambda 7605\AA$), are $\rm H_2O$ and $\rm O_2$ telluric lines and depressions.

\subsection{Input  parameters for model atmospheres  }
\label{inputparameters}
Using $m_v=6\fm71 $ from Table ~\ref{table2} and $\triangle m=0\fm18$ as the average of all 19 $\triangle m$ measurements (Table ~\ref{points}), since there is no significant  difference in $\triangle m$ under different filters between $\lambda$503-800 nm, we calculated a preliminary individual $m_v$ for each component as:  $m_{va}=7\fm376$ and $m_{vb}=7\fm556$.

These visual magnitudes  along with the system's parallax  from Hipparcos catalogue ($\pi=13.94\pm0.90,d=71.74 pc$) and the extinction coefficient $A_v=0.1432$ mag from \cite{2011ApJ...737..103S} and Galactic Dust Reddening and Extinction Archive (http://irsa.ipac.caltech.edu/applications/DUST/) with the following equation:
\begin{eqnarray}
\label{Mv}
M_v=m_v+5-5\log(d)-A,
\end{eqnarray}
\noindent
give the absolute individual magnitudes as:

$M_v^{a}=2\fm95$ and $M_v^{b}=3\fm13$.

So, the correspondent effective temperatures for such absolute magnitudes would be either $T_{\rm eff}^{a}=6950 \rm{K} \,\,\& \,  T_{\rm eff}^{b}=6800 \rm{K} $ in the case of main sequence components as given by the Tables of \cite{2005oasp.book.....G} or less in the case of evolved components. Hence, the gravity acceleration constant at the surface of such stars would be log$g\leq 4.3$.

These values of the effective temperatures and gravity accelerations represent the preliminary input parameters for atmospheric modeling of both components, from which we compute their synthetic spectra.

\subsection{Synthetic spectra}

The spectral energy distributions in the continuous spectrum for each component is computed depending on solar abundance model atmospheres of each component using  grids of  Kurucz's 1994 blanketed models (ATLAS9).

The total energy flux from a binary star is created from the net
luminosity of the components $a$ and $b$ located at a distance $d
$ from the Earth \citep{2012PASA...29..523A}. So we can write:
%$$
%   \textrm{F}_\lambda .d^2 =
\begin{eqnarray}
\label{F1}
   F_\lambda \cdot d^2 = H_\lambda ^a \cdot R_{a} ^2 + H_\lambda ^b
\cdot R_{b} ^2,
\end{eqnarray}
 \noindent from which

\begin{eqnarray}
\label{F2}
 F_\lambda  = (R_{a} ^2/d)^2(H_\lambda ^a + H_\lambda ^b \cdot(R_{b}/R_{a})^2) ,
\end{eqnarray}
\noindent
 where $H_\lambda ^a $ and  $H_\lambda ^b$ are the fluxes from a unit
surface of the corresponding component. $F_\lambda$ here
represents the entire SED of the system.

When we built  synthetic SEDs using the aforementioned preliminary input parameters  (Sec. \ref{inputparameters}) and equations ~\ref{F1} \& ~\ref{F2},  and compared them with the observational SED, we found that there was no coincidence between them within the criteria of the best fit which are  the maximum values of the absolute fluxes, the inclination of the continuum of the spectra, and the profiles of the absorption lines.

So, many attempts were made to achieve the best fit between the synthetic SEDs and the observational one using the iteration method of different sets of parameters according to the following equations:
\begin{eqnarray}
\label{RLT}
\log(R/R_\odot)= 0.5 \log(L/L_\odot)-2\log(T/T_\odot)\\
\log g = \log(M/M_\odot)- 2\log(R/R_\odot) + 4.43,
\end{eqnarray}
\noindent  where
$T_\odot=5777\rm{K}$ was used.
But in all attempts of modeling the components as main sequence stars, there were  disagreements between synthetic and observational SEDs in both; the inclination of the continuum (which represents the effective temperatures) and the absolute flux (which represents either the radii of the components or the parallax of the system).

The key parameters to get the best fit are the radii of the components, which should be bigger than what would be if they were main sequence stars (as proposed formerly), and the effective temperatures which should be lower. That means both components are evolved stars.

\begin{figure}[!ht]
\resizebox{\hsize}{!} {\includegraphics[]{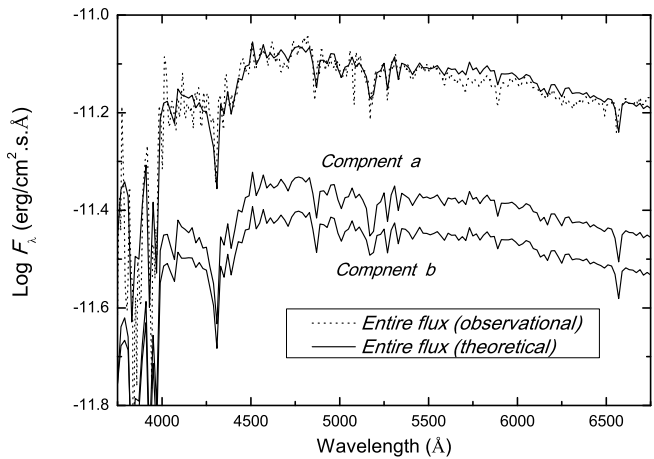}}
% \centerline{\psfig{figure=ModlFit1291a.eps,width=0.5\textwidth,clip=}}%,width=14cm}}
 \caption{Dotted line: the entire observational SED in the continuous spectrum of the
 system \citep{2002BSAO...53...58A}. Solid lines: the  entire computed SED of the two components,
  the computed flux of the primary component with $T_{\rm eff}=5625\pm75$\,K,
 log $g=3.75\pm0.25, R=2.69\pm0.30 R_\odot$, and the computed flux of the secondary
component with
 $T_{\rm eff} =5675\pm75$\,K, log $g=3.75\pm0.25$,  $R=2.42\pm0.30 R_\odot $ and $d=69.31$ pc.}
 \label{Modfit}
\end{figure}

So, depending on that proposal, hundreds of models were built and compared with the observational SED tell  the best fit was achieved using the following set of parameters (Fig. ~\ref{Modfit}):

$$ T_{\rm eff}^{a}
=5625\pm75{\rm K}, T_{\rm eff}^{b} =5675\pm75{\rm K},$$ $$ \log
g_{a}=3.75\pm0.25, \log g_{b}=3.75\pm0.25,$$
$$ R_{a}=2.69\pm0.30R_\odot,$$ and $$ R_{b}=2.42\pm0.30R_\odot.$$
%$d=24.80\pm3.20$\,pc.
\noindent
Thus the luminosities follow as:
$L_a=6.51\pm0.75 L_\odot$,  and $L_b=5.45\pm0.75 L_\odot$.

To  ensure the correlation between physical and geometrical parameters and to connect the dynamical analysis with the atmospheric modeling, we used Kepler's equation:
\begin{eqnarray}
\label{eq9}
\ (\frac{M_a + M_b}{M_\odot})(\pi^3)=\frac{a^3}{p^2},\\
\label{eq10}
\ \frac{\sigma_M}{M}=\sqrt{(3\frac{\sigma_\pi}{\pi})^2+(3\frac{\sigma_a}{a})^2+(2\frac{\sigma_p}{p})^2,}
\end{eqnarray}
\noindent where $(\frac{M_a + M_b}{M_\odot})$ is the mass sum of the two components in terms of solar mass, $\pi$ is the parallax of the system in arc seconds, $a$ is the semi-major axes in arc seconds and $p$ is the period in years.

Using the orbital elements of the system calculated in section \ref{orbital_elements} and $\pi=13.94\pm0.90$ as given by Hipparcos (Table ~\ref{table2}), the mass sum of the two components follows  as:
\begin{equation}
(M_a + M_b)/M_\odot = 3.06\pm 0.59,
\end{equation}
\noindent  which disagrees with the masses estimated from the evolutionary tracks based on atmospheres modeling. So, fixing the right part of  equation ~\ref{eq9} and changing the values of the left part in an iterated way,  ensures agreement between the masses of the components, their positions on the evolutionary tracks and the best fit between the synthetic and observational SED's.

For stars with formerly estimated physical and geometrical parameters,  this agreement is possible only for a system at a distance $d= 69.31$ pc ($\pi=14.43$mas).

The final physical and geometrical  parameters of the system are listed in Table \ref{tableFinala}, which represent the  parameters of the system's components within the error values of the measured quantities. Depending on the tables of \cite{2005oasp.book.....G} or  \cite{1992adps.book.....L},  the spectral types  of the  components are estimated as G6 and G5 for primary and secondary components respectively.

\section{Synthetic photometry}
In addition to the direct comparison, we can check reliability of our method of estimating the physical and geometrical parameters  by comparing the observed  magnitudes of the entire system from different ground or space based telescopes  with the entire synthetic ones. For that, we used the following relation \citep{{2006AJ....131.1184M},{2007ASPC..364..227M}}:
\begin{equation}
m_p[F_{\lambda,s}(\lambda)] = -2.5 \log \frac{\int P_{p}(\lambda)F_{\lambda,s}(\lambda)\lambda{\rm d}\lambda}{\int P_{p}(\lambda)F_{\lambda,r}(\lambda)\lambda{\rm d}\lambda}+ {\rm ZP}_p\,,
\end{equation}
to calculate the entire and individual synthetic magnitudes of the system, where $m_p$ is the synthetic magnitude of the passband $p$, $P_p(\lambda)$ is the dimensionless sensitivity function of the passband $p$, $F_{\lambda,s}(\lambda)$ is the synthetic SED of the object and $F_{\lambda,r}(\lambda)$ is the SED of the reference star (Vega).  Zero points (ZP$_p$) from  \cite{2007ASPC..364..227M} (and references there in) were adopted.

 The results of the calculated magnitudes and color  indices of the entire system and individual components, in different photometrical systems,  are shown in Table~\ref{synth2}.

\begin{table}[!ht]
\small
\begin{center}
\caption{ Magnitudes and color indices  of the synthetic spectra of the  system.}
\label{synth2}
\begin{tabular}{lcccc}
\noalign{\smallskip}
\hline
\noalign{\smallskip}
Sys. & Fil. & entire & comp. a & comp. b  \\
\hline
\noalign{\smallskip}
Joh-           & $U$ & $7.72\pm0.03$ & $8.41$ & $8.55$ \\
 Cou.          & $B$ & 7.45  &  8.13 &  8.28  \\
               & $V$ & 6.71 &  7.38 &  7.56 \\
               & $R$ & 6.32 &  6.99 & 7.18  \\
               &$U-B$& 0.28& 0.29 & 0.27 \\
               &$B-V$&0.73&  0.74 &  0.72 \\
               &$V-R$& 0.39&  0.40 & 0.38 \\
  \hline
\noalign{\smallskip}
  Str\"{o}m.        & $u$ & 8.88 & 9.55 &  9.72  \\
                    & $v$ & 7.84  & 8.52  &  8.66 \\
                    & $b$ & 7.12 & 7.79 &  7.97 \\
                    &  $y$& 6.68 & 7.35 &  7.53  \\
                    &$u-v$& 1.04 &1.03& 1.06 \\
                    &$v-b$& 0.72& 0.73 & 0.70 \\
                    &$b-y$& 0.44& 0.44& 0.44 \\
  \hline
\noalign{\smallskip}
  Tycho       &$B_T$  & 7.64 & 8.32 & 8.46   \\
              &$V_T$  & 6.79 &7.46 & 7.64  \\
              &$B_T-V_T$& 0.84& 0.86& 0.83\\
\hline
\noalign{\smallskip}
\end{tabular}
\end{center}
\end{table}

%\section {Formation and evolution of the systems}
\section {Results and discussion}

 A comparison between the  synthetic magnitudes and colors (Table~\ref{synth2}) with the observational ones (Tables~\ref{table2} and \ref{tableresult})  shows a very good consistency within the three photometrical systems Johnson-Cousins, Str\"{o}mgren and Tycho. Also,  synthetic  visual magnitudes of the two components (~\ref{synth2}) fit exactly those calculated in section ~\ref{inputparameters}.  This gives a good indication for the reliability of the method and the  estimated parameters of the individual components of the system, which are listed  in Table~\ref{tableFinala}.

 Moreover, there is a good consistency between the estimated absolute magnitudes as $M_v^{a}= 2\fm80\pm0.30, M_v^{b}=2\fm93\pm0.30$  with those previously calculated in section ~\ref{inputparameters} as $M_v^{a}=2\fm95, M_v^{b}=3\fm13$, and those given by \cite{2002aaa...385...87B} as $M_v^{a}=3.1 \& M_v^{b}=3.3$.

  Dynamical parallax of the system is introduced here as $d= 69.31$ pc ($\pi=14.43$mas), within the errors of Hipparcos parallax measurement, to insure the consistency between the physical and  geometrical parameters, and to explain the R-L-T relation of such VCBS.

\begin{figure}[!ht]
%\centerline{\psfig{figure=evol.eps,width=0.9\textwidth,clip=}}
\resizebox{\hsize}{!} {\includegraphics[]{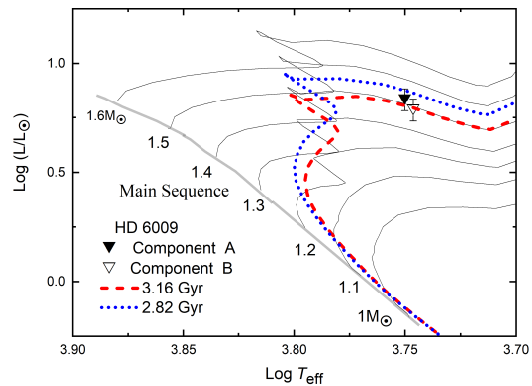}}
 \caption{The  system's components  on the evolutionary tracks of  \cite{2000A&AS..141..371G}. }
 \label{evol2}
\end{figure}

 Fig.~\ref{evol2} shows the positions of the components on the
 evolutionary tracks of  \cite{2000A&AS..141..371G}, where the error bars in the figure include the effect of the parallax  uncertainty.
 The age of the system can be established from the evolutionary tracks as 3.1 Gy.

 It is clear from the parameters of the system's components and their positions on the evolutionary tracks that they are  twin  evolved stars, with a bit bigger, more massive, more evolved and colder G6 IV primary  component. Hence, we can conclude, depending on the formation theories,  that  fragmentation is the most likely process for the formation of such system.
Where \cite{1994MNRAS.269..837B} concludes that fragmentation of rotating disk
around an incipient central protostar is possible, as long as
there is continuing infall, and \cite{2001IAUS..200.....Z} pointed out that
hierarchical  fragmentation during rotational collapse has been
invoked to produce binaries and multiple systems.

\begin{table}[!ht]
\small
\begin{center}
\caption{Parameters of the system's components.}
\label{tableFinala}
\begin{tabular}{l|c|c}
\noalign{\smallskip}
\hline
\noalign{\smallskip}
Component           & a                  &  b         \\
\hline
\noalign{\smallskip}
$T_{\rm eff}$\,(K)  & $5625\pm75$       & $5575\pm75$ \\
Radius (R$_{\odot}$)& $2.75\pm0.50$      & $2.65\pm0.50$ \\
$\log g$            & $3.75\pm0.25$      & $3.75\pm0.25$ \\
$L (L_\odot)$       & $6.80\pm0.75$     & $6.09\pm0.75$\\
$M_{v}$ $^\dag$     & $2\fm80\pm0.30$    & $2\fm93\pm0.30$\\
%$M_{bol}$ & $?\fm15\pm0.20$ & $?\fm38\pm0.25$\\
Mass, ($M_{\odot})$ & $1.42\pm0.15$      & $1.40 \pm0.15$  \\
$\overline{\rho}(\overline{\rho}_\odot)$& $0.07\pm0.02$& $0.10\pm0.02$\\
Sp. Type$^\ddag$    & G6                &G6 \\
luminosity class    & IV                  & IV      \\
\hline
%\multicolumn{1}{l}{Parallax (mas) }
%    & \multicolumn{2}{|c}{$40.32 \pm 5.00 $}\\
  \multicolumn{1}{l}{Age (Gy) }
    & \multicolumn{2}{|c}{ $3.1\pm 0.5$}\\
\hline
\noalign{\smallskip}
\end{tabular}
\end{center}
$^\dag${depending on the individual synthetic spectra (Table ~\ref{synth2})}.\\
$^\ddag${depending on the tables of  \cite{2005oasp.book.....G}}.
\end{table}

\begin{table}[!ht]
\small
\begin{center}
\caption{Comparison between the observational and synthetic  entire magnitudes, colours and magnitude differences of the system.}
\label{tablecopmarison}
\begin{tabular}{lcc}
%\noalign{\smallskip}
\hline \hline
%\noalign{\smallskip}
              & $\textrm{Obs}.^*$    & Synth. (this work)  \\
\hline
\noalign{\smallskip}

  $V_J$     & $6\fm71$          & $6\fm71\pm0.03$  \\
  $B_T$     & $7\fm683\pm0.008$ & $7\fm64\pm0.03$ \\
  $V_T$     & $6\fm806\pm0.006$ & $6\fm79\pm0.03$ \\
  $(B-V)_J$ & $0\fm783\pm0.007$ & $0\fm84\pm0.04$ \\
  $\triangle m $& $0\fm18^\dag$ & $0\fm18\pm0.04$\\

\hline \hline
%\noalign{\smallskip}
$^*${See Table~\ref{table2}}.\\
$^\dag${See section ~\ref{inputparameters} }.
\end{tabular}
\end{center}
\end{table}

\section{Conclusions}

 On the basis of analyzing  the binary system HD6009 using the method of  atmospheric modeling and dynamical analysis  for studying VCBSs, the following main conclusions can be drawn:

\begin{enumerate}
    \item The complete set of the physical and geometrical  parameters of the system's  components were estimated
  depending on the best fit between the observational SED  and
  synthetic ones built using the atmospheric modeling of the individual components.
  \item A modification to the parallax of the system is introduced within the error of Hipparcos parallax measurement.
  \item The estimated parameters show a good consistency with the previously published ones.
   \item Depending on the parameters of the system's components and their positions on the evolutionary tracks, we showed  that the system consists of a twin   G6  and G5 subgiant stars.
   \item The entire and individual $UBVR$ Johnson-Cousins, $uvby$ Str\"{o}mgren and $BV$ Tycho synthetic magnitudes and colors of the system were calculated.
   \item Finally, the fragmentation was proposed as the most likely process for the formation and evolution of the system.
\end{enumerate}

%\begin{acknowledgements}
% This work made use of CHORIZOS code of photometric and spectrophotometric data  analysis %(http://www.stsci.edu/$\sim$jmaiz/ software/chorizos/chorizos.html).
%\end{acknowledgements}
\section*{Acknowledgments}
The author would like to thank Mrs. Kawther Al-Waqfi for her help in some calculations. A part of this work was done during the research visit of the author to Max Plank Institute for Astrophysics-Garching in 2011 which was funded by The Deutsche Forschungsgemeinschaft (DFG, German Research Foundation). This work made use of the Fourth Interferometric Catalogue, SIMBAD database, Astrophysics Data System Bibliographic Services (NASA), IPAC data systems and CHORIZOS code of photometric and spectrophotometric data  analysis (http: //www.stsci.edu/ jmaiz/software/ chorizos/chorizos.html) under the  Interactive Data Language IDL-ITT Visual Information Solutions.

%\bibliographystyle{raa}
%\bibliographystyle{elsarticle-harv}
%\bibliographystyle{model2-names}
%\bibliography{alwardat}

\end{document}